\documentclass[prl,aps,amsmath,amssymb,twocolumn,groupedaddress,floats,final,postprint,superscriptaddress]{revtex4}
\usepackage{graphicx}

\usepackage[polutonikogreek,english]{babel}
\usepackage{teubner}
\usepackage{color}
\usepackage{bm}
\usepackage{MnSymbol}%
\usepackage{wasysym}%
\usepackage{algorithm}
\usepackage{hyperref}
\usepackage{float}
\usepackage{pgf}
\usepackage{tikz}
\usepackage{graphicx}
\usepackage{subfigure}

\usepackage[mathlines]{lineno}

\newcommand{\ZZ}{\mathbb{Z}}

\begin{document}

\author{Zongzheng Zhou}
\affiliation{ARC Centre of Excellence for Mathematical and Statistical Frontiers
(ACEMS), School of Mathematical Sciences, Monash University, Clayton,
Victoria 3800, Australia}
\author{Jens Grimm}
\email{jens.grimm@monash.edu}
\affiliation{ARC Centre of Excellence for Mathematical and Statistical Frontiers
(ACEMS), School of Mathematical Sciences, Monash University, Clayton,
Victoria 3800, Australia}
\author{Sheng Fang}
\affiliation{Department of Modern Physics, University of Science and Technology of China, Hefei, Anhui 230026, China}
\author{Youjin Deng}
\email{yjdeng@ustc.edu.cn}
\affiliation{Department of Modern Physics, University of Science and Technology of China, Hefei, Anhui 230026, China}
\affiliation{National Laboratory for Physical Sciences at Microscale, University of Science and Technology of China, Hefei, Anhui 230026, China}
\author{Timothy M. Garoni}
\email{tim.garoni@monash.edu}
\affiliation{ARC Centre of Excellence for Mathematical and Statistical Frontiers
(ACEMS), School of Mathematical Sciences, Monash University, Clayton,
Victoria 3800, Australia}

\title{Random-length Random Walks and Finite-size Scaling in high dimensions}

\date{\today}

\begin{abstract} 
We address a long-standing debate regarding the finite-size scaling of the Ising model in high dimensions, by introducing a
\emph{random-length random walk} model, which we then study rigorously. We prove that this model exhibits the same universal FSS behaviour
previously conjectured for the self-avoiding walk and Ising model on finite boxes in high-dimensional lattices. Our results show that the
mean walk length of the random walk model controls the scaling behaviour of the corresponding Green's function. We numerically demonstrate
the universality of our rigorous findings by extensive Monte Carlo simulations of the Ising model and self-avoiding walk on five-dimensional
hypercubic lattices with free and periodic boundaries.

\end{abstract}

\maketitle

Finite-size Scaling (FSS)~\cite{fisher_1971,fisher_1972} is a fundamental theory which characterizes the asymptotic approach of finite
systems to the thermodynamic limit, close to a continuous phase transition. While critical systems above the upper critical dimension
$d_{\rm c}$ exhibit simple mean-field behaviour in the thermodynamic limit~\cite{fernandez_1992}, their FSS behaviour above $d_{\rm c}$ is
surprisingly subtle and the subject of long-standing debate; see e.g.~\cite{lundow_2016, flores-sola_2016,
  grimm_2017,wittmann_2014,kenna_2014,berche_2012,lundow_2014}. In this work, we clarify a number of these subtleties by introducing a
simple model, which can be studied rigorously.

The $n$-vector model~\cite{stanley_1968}, which describes interacting spin systems on a lattice, plays a central role in various areas of
physics such as statistical mechanics and condensed matter physics. Prominent examples are the Self-avoiding Walk (SAW) ($n \to 0$) in
polymer physics, and the Ising ($n=1$) and XY ($n=2$) models of ferromagnetism. The latter can be related to the Bose-Hubbard
model~\cite{svistunov_2015} which describes bosonic atoms in an optical lattice.

On an infinite hypercubic lattice $\mathbb{Z}^d$, it is known rigorously~\cite{hara_2008,sakai_2007} that for sufficiently large dimension
$d$, the two-point functions of the critical Ising and SAW models exhibit the same scaling behaviour as the Green's function of the Simple
Random Walk (SRW). On finite lattices this connection breaks down because SRW is recurrent, implying that its Green's function does not
exist.

In this Letter, we argue that if one considers random walks with an appropriate random (finite) length $\mathcal{N}$, then the Green's
function displays the same finite-size scaling as the two-point functions of the SAW and Ising models, defined on boxes in $\mathbb{Z}^d$ of
linear size $L$. For this Random-length Random Walk (RLRW) model, one can prove~\cite{zhou_2018} that if $d \ge 3$ and $\langle \mathcal{N}
\rangle \asymp L^\mu$ with $\mu \ge 2$, then the Green's function scales as
\begin{equation}
g(\mathbf{x}) \asymp \begin{cases} \|\mathbf{x}\|^{2-d}, & \|\mathbf{x}\| \le O\left(L^{(d-\mu)/(d-2)}\right)\\ 
L^{\mu-d},  & \|\mathbf{x}\| \ge  O\left(L^{(d-\mu)/(d-2)}\right).
\end{cases}
\label{eq:prel_theorem_green}
\end{equation}

In words, if $\mu>2$, $g(\mathbf{x})$ exhibits the standard infinite-lattice asymptotic decay $\|\mathbf{x}\|^{2-d}$ at moderate values of
$\mathbf{x}$, but then enters a plateau of order $L^{\mu-d}$ which persists to the boundary.  Since a typical RLRW will explore distances of
order $\sqrt{\langle \mathcal{N} \rangle}$ from the origin, no plateau exists for $\mu <2$ because typical walks will be too short to feel
the boundary; in this case $g(\mathbf{x})$ decays significantly faster~\cite{zhou_2018} than $\|\mathbf{x}\|^{2-d}$ for $\|\mathbf{x}\|\gg
\sqrt{\langle\mathcal{N}\rangle}$.

The above scaling behaviour of the Green's function holds on boxes with both free and periodic boundaries.
As a consequence of this scaling~\cite{footnote:sus}, one can prove~\cite{zhou_2018} that the corresponding susceptibility scales as
\begin{equation}
\chi \asymp L^\mu, \qquad \text{for any $\mu>0$.}
\label{eq:prel_theorem_sus}
\end{equation}

The mean walk length of SAW, restricted to a finite box in $\mathbb{Z}^d$, depends strongly on the boundary conditions imposed. For a given
choice of SAW boundary conditions, one can consider a RLRW where $\langle \mathcal{N} \rangle$ is chosen to scale in the same way as it does
for the SAW. Our numerical results below strongly suggest that the scaling of the Green's function of this RLRW model, given by Eq.~\eqref{eq:prel_theorem_green}, then correctly predicts the two-point function scaling of the corresponding SAW model. We therefore
conclude that the SAW two-point function is only affected by geometry via its effect on the mean walk length. These observations are seen to
hold not only at the thermodynamic critical point, but also at general pseudo-critical points. We numerically demonstrate the universality
of these predictions by showing that they also correctly describe the FSS behaviour of the Ising two-point function.

These observations shed light on a number of open questions regarding the FSS behaviour of the Ising model above $d_{\rm c}$. For periodic
boundary conditions (PBC) at criticality, the scaling of the Ising two-point function has been actively debated
in~\cite{kenna_2014,wittmann_2014,grimm_2017}. The known~\cite{yadin_2016} behaviour of the mean walk length of SAW on the complete graph~\cite{footnote_complete_graph},together with extensive Monte Carlo simulations in five dimensions, suggest that on high-dimensional tori at criticality we have $\langle
\mathcal{N} \rangle_{\rm SAW} \asymp L^{d/2}$.  We therefore predict that the critical SAW and Ising two-point functions should be given by Eq.~\eqref{eq:prel_theorem_green} with $\mu =d/2$. This prediction is in agreement with the conjectured behaviour of the critical Ising
two-point function given in~\cite{papathanakos_2006}, and is in excellent agreement with the numerical results presented
in~\cite{grimm_2017}.

For free boundary conditions (FBC), the possible existence of the FSS behaviour $\chi \asymp L^{d/2}$ at \textit{pseudo-critical} points is
the subject of ongoing debate~\cite{berche_2012,wittmann_2014,lundow_2016}. Specifically, denoting by $T_L$ the temperature which maximizes
$\chi(T,L)$ on a box of size $L$, it was observed numerically in~\cite{berche_2012} that $\chi(T_L,L)$ has the same $L^{d/2}$ scaling
observed at criticality for periodic systems. The results in~\cite{wittmann_2014} are in agreement with this observation, however, the more
recent work~\cite{lundow_2016} refuted this claim, and numerically observed only the standard mean-field scaling $L^2$. From Eq.~\eqref{eq:prel_theorem_sus}, we see that one \textit{can} observe $\chi \asymp L^{d/2}$ in a RLRW model in which the mean walk length
scales as $L^{d/2}$. Universality then suggests that this scaling should also be observable in SAW and Ising models, at appropriate
pseudo-critical points. Our numerical results below confirm this.

\begin{figure}
\includegraphics[scale=0.44]{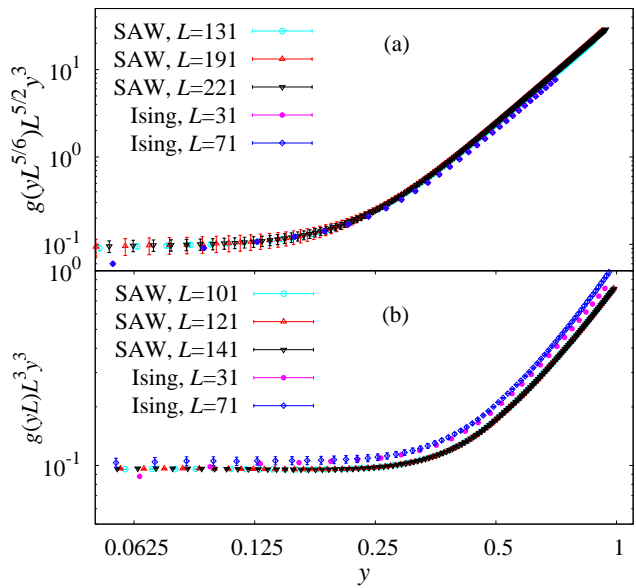}
\caption{Appropriate scaled two-point functions of the Ising model and SAW on five-dimensional hypercubic lattices with \textit{periodic} boundaries onto the scaling variable $y=\| \mathbf{x} \|/L^{(d-\mu)/(d-2)}$. (a) Anomalous FSS scaling at $z_{\rm c}$ onto the ansatz in Eq.~\eqref{eq:prel_theorem_green} with $\mu =d/2$. When $\| \mathbf{x} \| \approx L/2$, the two-point functions display the anomalous FSS behaviour $g(\mathbf{x}) \asymp L^{-d/2}$, in contrast to the standard mean field prediction $g(\mathbf{x}) \asymp L^{2-d}$. (b) Standard mean-field scaling at the pseudo-critical point $z_L = z_{\rm c}-aL^{-2}$ onto the ansatz in Eq.~\eqref{eq:prel_theorem_green} with $\mu = 2$. In contrast to the critical PBC case, the two-point functions display the standard mean-field scaling behaviour $g(\mathbf{x}) \asymp L^{2-d}$ when $\| \mathbf{x} \| \approx L/2$.}
\label{fig:correlation_pbc}
\end{figure}\medskip

\textit{Random-length Random Walk.---} Let $(S_t)_{t \in \mathbb{N}}$ be a simple random walk on a box of side length $L$ in $\ZZ^d$,
centered at the origin. Let $\mathcal{N}$ be an $\mathbb{N}$-valued random variable, independent of each choice of step in
$(S_t)_{t\in\mathbb{N}}$. We refer to $(S_t)_{t=0}^\mathcal{N}$ as the corresponding RLRW. We study its Green's
function
$$
g_{\rm RLRW}(\mathbf{x}): = \mathbb{E}\left(\sum_{n=0}^\mathcal{N} P(S_n=\mathbf{x})\right),
$$
which is the expected number of visits to $\mathbf{x}$, and the corresponding susceptibility $\chi_{\rm RLRW}:=\sum_{\mathbf{x}} g_{\rm
  RLRW}(\mathbf{x})$. Here, $P(S_n=\mathbf{x})$ denotes the probability that the RLRW is at site $\mathbf{x}$ after $n$ steps.

Consider a RLRW with mean walk length $N:=\langle \mathcal{N} \rangle \asymp L^\mu$ on a $d \ge 3$ dimensional
hypercubic lattice, with either periodic or free boundary conditions. If $\mu \ge 2$, it can then be proved~\cite{zhou_2018} that the Green's
function exhibits the piecewise asymptotic behaviour in Eq.~\eqref{eq:prel_theorem_green}. In particular, the case $\mu>2$ shows the existence of a macroscopic plateau of order $L^{\mu-d}$ for large distances, while this plateau is absent for $0<\mu<2$. The case $\mu=2$ is marginal. \medskip

\textit{Numerical setup for $n$-vector models.---}We study the two-point function $g_{\rm Ising}(\mathbf{x}) := \mathbb{E}(s_0
s_\mathbf{x})$ for the zero-field ferromagnetic Ising model, defined by the Hamiltonian $\mathcal{H} = - \sum_{ij}
\mathbf{s}_i\mathbf{s}_j$. Here, $\mathbf{s}_i=\pm 1$ denotes the spin at site $i$ of a hypercubic lattice of side length $L$,
and the sum is over nearest neighbours. We simulate the Ising model at fugacities $z:=\tanh(\beta)$, where $\beta$ is the inverse Ising
temperature, via the worm algorithm introduced in~\cite{prokofev_2001}.

We also investigate the SAW on a box with linear size $L$ in the variable length ensemble. We study the two-point function
$g_{\text{SAW}}(\mathbf{x}):=\sum_{\omega\,:\,\ 0 \to \mathbf{x}} z^{|\omega|}$, where the sum is over all SAWs starting at the origin $0$
and ending at $\mathbf{x}$. We simulated this ensemble using an irreversible version of the Berretti-Sokal
algorithm~\cite{berretti_1985,hu_2017}. For both models we study the corresponding susceptibility, defined by $\chi_{\rm Ising/SAW} : =
\sum_{\mathbf{x}} g_{\rm Ising/SAW}(\mathbf{x})$.

\begin{figure}
\includegraphics[scale=0.44]{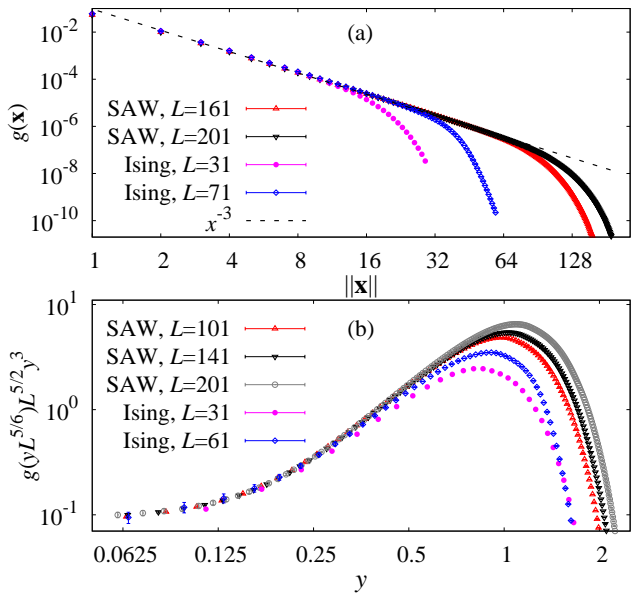}
\caption{Two-point functions of the Ising model and SAW on five-dimensional hypercubic lattices with \textit{free} boundaries. (a) Standard mean-field scaling $g(\mathbf{x}) \asymp \| \mathbf{x} \|^{2-d}$ at $z_{\rm c}$. (b) Anomalous FSS at the pseudo-critical point $\tilde{z}_L = z_{\rm c}+a_LL^{-2}$ onto the scaling variable $y=\| \mathbf{x} \|/L^{(d-\mu)/(d-2)}$ with $\mu=d/2$. The two-point functions collapse except at distances close to the boundary. This shows that $g(\mathbf{x})$ displays the same FSS behaviour as on periodic boundaries at criticality.}
\label{fig:correlation_free}
\end{figure}

We study our models on hypercubic lattices, in the case of both free and periodic boundary conditions. The Ising model was simulated at the
estimated location of the infinite-volume critical point $z_{\text{c,Ising,5d}} = 0.113~915~0(5)$~\cite{lundow_2014} in five dimensions, and
the simulations for the SAW were performed at the estimated infinite-volume critical point $z_{\text{c,SAW,5d}} =
0.113~140~84(1)$~\cite{hu_2017}. We also simulated the FSS behaviour at pseudo-critical points $z_L = z_{\rm c} - aL^{-\lambda}$ for various
$a \in \mathbb{R}$ and $\lambda>0$. We simulated linear system sizes up to $L=71$ in the Ising model and $L=201$ for the SAW. To estimate
the exponent value for a generic observable $Y$ we performed least-squares fits to the ansatz $Y = a_Y L^{b_Y} + c_Y$. A detailed analysis
of autocorrelation times can be found in~\cite{deng_2007} for the worm algorithm and in~\cite{hu_2017} for the irreversible Berretti-Sokal algorithm.\medskip

\textit{Universal scaling at criticality.---} We now argue that Eqs.~\eqref{eq:prel_theorem_green} and~\eqref{eq:prel_theorem_sus} correctly
predict the FSS behaviour of the two-point functions and the susceptibility of the critical SAW and Ising model, with either FBC or PBC.

We first study the periodic case. It is expected that models on high-dimensional tori should exhibit the same scaling as the corresponding
model on the complete graph. It was proved in~\cite{yadin_2016} that, on the complete graph, $N_{\rm SAW}$ scales at criticality like the
square root of the number of vertices. On five-dimensional tori, our fits for $N_{\rm SAW}$ at criticality lead to the exponent value
$2.50(1)$, in excellent agreement with the complete graph prediction of $d/2$.  Combining this scaling for $N_{\rm SAW}$ with our results
for the RLRW, the two-point functions of the critical Ising and SAW models on high-dimensional tori are then predicted to display the
scaling in Eq.~\eqref{eq:prel_theorem_green} with $\mu =d/2$. Figure~\ref{fig:correlation_pbc}(a) verifies this prediction, showing an
excellent data collapse for appropriately scaled versions of the two-point functions of the Ising and SAW models onto the scaling variable
$y:=\| \mathbf{x} \|/L^{(d-\mu)/(d-2)}$ with $\mu=d/2$. As a corollary of this two-point function scaling, we obtain $\chi \asymp L^{d/2}$,
in agreement with the numerical studies for the Ising model in~\cite{binder_1985,parisi_1996}, and with our direct exponent estimates for $d=5$
of $2.50(1)$ for $\chi_{\rm SAW}$, and $2.51(2)$ for $\chi_{\rm Ising}$.

On free boundaries at criticality, our fits for $N_{\rm SAW}$ lead to the exponent value $2.00(1)$, strongly suggesting that $N_{\rm SAW}
\asymp L^2$. Combining this scaling for $N_{\rm SAW}$ with our results for the RLRW, the two-point functions of the critical Ising and SAW
models on high-dimensional boxes with free boundaries are then predicted to display the scaling in Eq.~\eqref{eq:prel_theorem_green} with
$\mu = 2$. Figure~\ref{fig:correlation_free}(a) verifies this prediction, showing an excellent data collapse for the two-point
functions of the critical Ising and SAW models onto the ansatz in Eq.~\eqref{eq:prel_theorem_green} with $\mu =
2$. Equation~\ref{eq:prel_theorem_sus} then predicts $\chi \asymp L^2$, in agreement with the numerical study of the Ising model
in~\cite{lundow_2014}, and with our direct exponent estimates for $d=5$ of $2.01(8)$ for the Ising model and $1.99(1)$ for the SAW.\medskip

\textit{Universal scaling at pseudo-critical points.---} We now turn to the actively debated
question~\cite{berche_2012,wittmann_2014,lundow_2016} of whether one can observe the scaling behaviour $\chi \asymp L^{d/2}$, corresponding
to critical PBC behaviour, on free boundaries at pseudo-critical points. This also motivates the reverse question, of whether it is possible
to observe the standard mean-field behaviour $\chi \asymp L^2$, corresponding to critical FBC behaviour, at pseudo-critical points on
periodic boundaries. The above results for the RLRW suggest that the FSS behaviour of the SAW two-point function should only
depend on the boundary conditions through their effect on $N$. We now numerically verify that this is indeed the case, and that 
analogous results also hold for the Ising model.

For periodic boundaries, we study FSS at pseudo-critical points $z_L(\lambda) = z_{\rm c} - aL^{-\lambda}$, with $a$ chosen positive so that the
walk lengths are \emph{decreased} compared with criticality. On the complete graph, it can be shown~\cite{zhou_2018} that at a
pseudo-critical point $z_V(\zeta) = z_{\rm c} - a V^{-\zeta}$ we have $N_{\rm SAW} \asymp V^{1/2}$ if $\zeta \ge 1/2$, while $N_{\rm SAW}
\asymp V^{\zeta}$ if $\zeta \le 1/2$.  Considering a RLRW on a high-dimensional torus, whose mean walk length scales in this way, the Green's function and
susceptibility then scale as in Eqs.~\eqref{eq:prel_theorem_green} and~\eqref{eq:prel_theorem_sus} with $\mu = \zeta d=:\lambda$ for any $0< \lambda\le
d/2$, and $\mu=d/2$ for $\lambda\ge d/2$. By universality, we then expect the same behaviour to hold for both SAW and the Ising model at the
pseudo-critical point $z_L(\lambda)$ on high-dimensional tori.

Taking $\lambda=2$, the above argument predicts that the pseudo-critical two-point functions display the mean-field behaviour
$g(\mathbf{x}) \asymp \| \mathbf{x} \|^{2-d}$. Fig.~\ref{fig:correlation_pbc}(b) shows an appropriately scaled version of the two-point
functions of the Ising model and SAW onto the ansatz in Eq.~\eqref{eq:prel_theorem_green} with $\mu=2$. The excellent data collapse provides
strong evidence for the predicted existence of standard mean-field behaviour at $z_L(2)$.

We emphasize that, despite appearances, the two-point functions in Fig.~\ref{fig:correlation_pbc}(a) and (b)
do not display the same FSS behaviour. In particular, it follows from the scaling ansatz in Eq.~\eqref{eq:prel_theorem_green} that if $\| \mathbf{x} \| \approx L/2$, then the critical two-point
functions scale as $g(\mathbf{x}) \asymp L^{-d/2}$, while $g(\mathbf{x}) \asymp L^{2-d}$ at $z_L(2)$.

\begin{figure}
\includegraphics[scale=0.44]{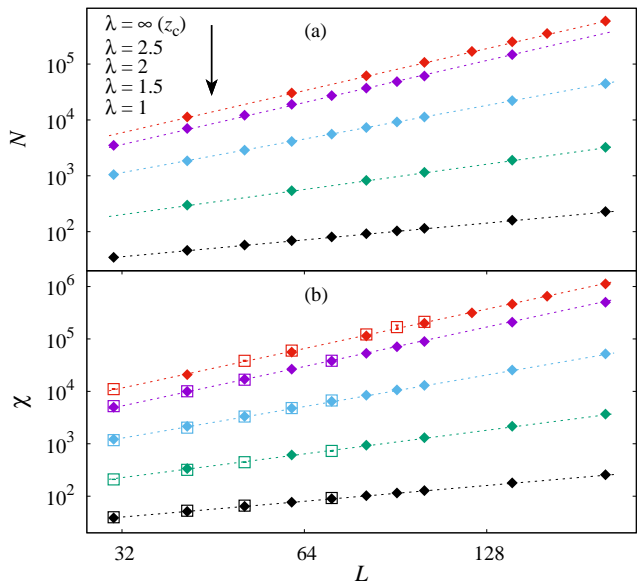}
\caption{FSS behaviour of the mean walk length $N_{\rm SAW}$ (top) and the susceptibility $\chi_{\rm Ising, SAW}$ (bottom) with
  \textit{periodic} boundary conditions in five dimensions. The diamonds (squares) display SAW (Ising) data. To emphasize universality, the
  Ising and SAW data were translated onto the same curve. In each figure, the line on the top corresponds to the critical scaling behaviour
  $N,\chi \asymp L^{d/2}$. The remaining lines (top to bottom) correspond to the scaling behaviour $N,\chi \asymp L^\lambda$ at
  pseudo-critical points $z_L(\lambda)=z_{\rm c}-aL^\lambda$ ($a>0$) with $\lambda= 2.5,~2,~1.5,~1$. }
\label{fig:walk_sus_pbc}
\end{figure}

Considering more general values of $\lambda$, Fig.~\ref{fig:walk_sus_pbc}(a) shows the scaling of $N_{\rm SAW}$ at $z_L(\lambda)$ on
five-dimensional tori for $\lambda = 1,~1.5,~2,~2.5$. Our fits lead to the exponent values $0.998(2)$
for $\lambda = 1$, $1.499(2)$ for $\lambda = 1.5$, $2.01(1)$ for $\lambda = 2$, and $2.47(4)$ for $\lambda = 2.5$, in excellent agreement
with the corresponding results on the complete graph.  Figure~\ref{fig:walk_sus_pbc}(b) then shows the scaling behaviour of the
susceptibility for $\lambda =1,~1.5,~2,~2.5$. Our fits for the SAW lead to the exponent values $1.005(6)$ for $\lambda = 1$, $1.503(5)$ for
$\lambda = 1.5$, $2.00(1)$ for $\lambda = 2$, $2.46(5)$ for $\lambda = 2.5$. For the Ising model, our fits lead to $1.00(1)$ for $\lambda =
1$, $1.51(2)$ for $\lambda = 1.5$, $2.05(7)$ for $\lambda = 2$, and $2.4(1)$ for $\lambda = 2.5$. These estimates are all in excellent agreement with above predictions.

\begin{figure}
\includegraphics[scale=0.44]{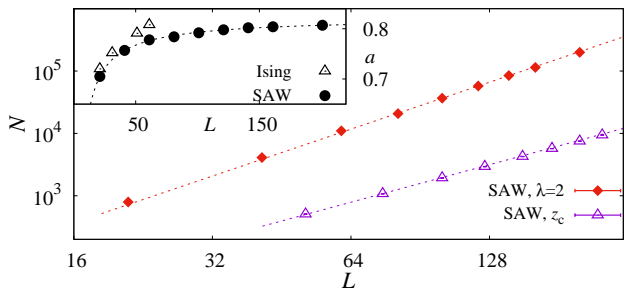}
\caption{FSS behaviour of the mean walk length $N_{\rm SAW}$ with \textit{free} boundary
  conditions in five dimensions. The inset shows the convergence of $a_L$ in the SAW (circles) and Ising model (triangles).}
\label{fig:walk_sus_fbc}
\end{figure}

Finally, we consider pseudo-critical behaviour with free boundary conditions. There has been considerable
debate~\cite{berche_2012,wittmann_2014,lundow_2016} concerning the existence of critical PBC FSS behaviour on lattices with FBC at a
pseudo-critical point which maximizes $\chi(T,L)$ on a box of linear size $L$. It has been numerically established that this pseudo-critical
point has shift exponent $\lambda =2$~\cite{berche_2012,wittmann_2014,lundow_2016}.  A simple methodology to gauge the possibility of
observing $\chi \asymp L^{d/2}$ at such a pseudo-critical point is to define a sequence $a_L$ such that
$\chi_{\text{FBC},\tilde{z}_L}(L)=\chi_{\text{PBC},z_{\rm c}}(L)$ with $\tilde{z}_L:=z_{\rm c} + a_L L^{-2}$, and to then show that $a_L$
converges.  If such a convergent sequence exists, this approach forces $\chi_{\text{FBC},z_L}$ to scale as $L^{d/2}$, where $z_L=z_{\rm c} +
a_{\infty}L^{-2}$. The inset of Fig.~\ref{fig:walk_sus_fbc} shows the sequence $a_L$ in the Ising and SAW models.  For SAW, the series $a_L$
clearly appears to converge, and our fits predict $a_{\text{SAW},\infty} = 0.824(2)$.  The Ising data are roughly consistent with the SAW
data, albeit over a much smaller range of $L$ values.

Fitting the FBC data for $N_{\rm SAW}$ at $\tilde{z}_L$ produces an exponent estimate of $2.48(6)$, suggesting that $N_{\rm SAW} \asymp
L^{d/2}$, compared with $N_{\rm SAW}\asymp L^2$ at $z_{\rm c}$; see Fig.~\ref{fig:walk_sus_fbc}.  Universality then suggests that the Ising
and SAW two-point functions should follow Eq.~\eqref{eq:prel_theorem_green} with $\mu=d/2$. Figure~\ref{fig:correlation_free}(b) shows the
appropriately re-scaled two-point functions.  We observe excellent data collapse, except at distances close to the boundary. This strong
boundary effect may explain the apparent discrepancies~\cite{berche_2012,wittmann_2014,lundow_2016} in determining the correct scaling
behaviour for the pseudocritical Ising model with FBC. Regardless, we conclude from Fig.~\ref{fig:correlation_free}(b) that the anomalous
FSS behaviour, observed on periodic boundaries at criticality, \emph{can} be observed on free boundaries, in agreement
with~\cite{berche_2012,wittmann_2014}.\medskip

\textit{Discussion.---} In this Letter, we have introduced a random-length random walk model to clarify a number of open questions regarding
the FSS behaviour of the Ising model above $d_{\rm c}$. For periodic boundaries, by combining the RLRW model with the scaling of the mean
walk length of SAW on the complete graph, we were able to predict the asymptotic scaling of the Ising and SAW two-point functions on
high-dimensional tori at a family of pseudo-critical points $z_L(\lambda)=z_{\rm c}-aL^{-\lambda}$, and showed
that the scaling exponents vary continuously with $\lambda$ when $0<\lambda \le d/2$. As special cases, at $z_{\rm c}$ we recovered the
behaviour conjectured in~\cite{papathanakos_2006}, while at $z_L(2)$ we showed the Ising two-point function displays standard mean-field
behaviour.

On free boundaries, combining the RLRW model with the numerical scaling of $N_{\rm SAW}$ predicts that the critical Ising two-point function
displays standard mean-field decay. It follows that the susceptibility scales as $L^2$, in agreement with the numerical observation
in~\cite{lundow_2014}. We also studied the actively debated FSS behaviour at the pseudo-critical point $z_L = z_{\rm c}+aL^{-2}$. We
established that the Ising two-point function displays the same FSS behaviour as on periodic boundaries at criticality, in agreement with
the numerical observations in~\cite{berche_2012,wittmann_2014}.

Recently, three-dimensional quantum spin models, which are related to the corresponding four-dimensional classical
counterpart~\cite{footnote_mapping}, have been the subject of intensive theoretical, experimental and numerical
studies~\cite{qin_2017,bissbort_2011, ruegg_2008}. Our work has focused on the FSS behaviour of the $n$-vector model above $d_{\rm
  c}=4$. Although \textit{at}~$d_{\rm c}$ the situation is likely complicated by logarithmic corrections, we believe that our results for
$d>d_{\rm c}$ are a necessary first step in understanding the correct scaling behaviour for applications to three-dimensional quantum spin
models.\medskip

\begin{acknowledgments}
We would like to thank Andrea Collevecchio, Eren M. El\c{c}i and Kevin Leckey for fruitful discussions. This work was supported under the Australian Research
Council's Discovery Projects funding scheme (Project Number DP140100559). It was undertaken with the assistance of resources from the
National Computational Infrastructure (NCI), which is supported by the Australian Government. This research was supported in part by the
Monash eResearch Centre and eSolutions-Research Support Services through the use of the MonARCH HPC Cluster. J. Grimm acknowledges Monash University and the Australian Mathematical Society for their financial support. Y. Deng and S. Fang acknowledges the support by the
National Key R\&D Program of China under Grant No.~2016YFA0301604 and by the National Natural Science Foundation of China under Grant No.~ 11625522.
\end{acknowledgments}


\begin{thebibliography}{99}
\bibitem{footnote:sus} To guarantee $g(\mathbf{x})$ decays sufficiently fast at large $\|\mathbf{x}\|$ for~\eqref{eq:prel_theorem_sus} to
  hold when $\mu<2$, we make the physically reasonable assumption $\sqrt{\text{Var}(\mathcal{N})} = O(\langle \mathcal{N} \rangle)$.
\bibitem{footnote_complete_graph} The complete graph on $V$ vertices is a graph in which each pair
of vertices is adjacent, so that there are $V(V-1)/2$ edges in total.
\bibitem{footnote_mapping} More generally, the quantum system in $d$ spatial dimensions is related to the
classical model in $d+1$ dimensions.
\end{thebibliography}
\end{document}